\documentclass[twocolumn,showpacs,amsmath,prl,aps,amssymb]{revtex4-1} 

\usepackage[T1]{fontenc}
\usepackage[utf8]{inputenc}
\usepackage{times}
\usepackage{amssymb,amsmath}
\usepackage{amsbsy}
\usepackage[pdftex]{graphicx}
\usepackage{bm}
\usepackage{float}

\usepackage[unicode,breaklinks]{hyperref}
\hypersetup{
    unicode=true,
    plainpages=false, 
    colorlinks=true,
    linkcolor=blue,
    citecolor=blue,
    filecolor=black,
    urlcolor=blue
}
\usepackage{url}
\usepackage{verbatim}
\usepackage{siunitx}
\newcommand{\add}{a_{dd}}
\newcommand{\edd}{\epsilon_{dd}}
\newcommand{\gdd}{g_{dd}}
\newcommand{\br}{\mathbf{r}}
\newcommand{\bx}{\mathbf{x}}
\newcommand{\bd}{\mathbf{d}}
\newcommand{\bdb}{{\bar{\mathbf{d}}}}
\newcommand{\bsigma}{\boldsymbol{\sigma}}

\newcommand{\Eb}{\bar{E}}

\newcommand{\var}{\mathrm{var}}
\newcommand{\LGP}{\mathcal{L}_\mathrm{GP}}
\newcommand{\gammaQF}{\gamma_{\mathrm{QF}}}

\begin{document}

\title{Droplet crystal ground states of a dipolar Bose gas}

\author{D.~Baillie}  
\affiliation{Dodd-Walls Centre for Photonic and Quantum Technologies, New Zealand}
\affiliation{Department of Physics, University of Otago, Dunedin 9016, New Zealand}
\author{P.~B.~Blakie}   
\affiliation{Dodd-Walls Centre for Photonic and Quantum Technologies, New Zealand}
\affiliation{Department of Physics, University of Otago, Dunedin 9016, New Zealand}

\begin{abstract}  
We show that the ground state of a dipolar Bose gas in a cylindrically symmetric harmonic trap has a rich phase diagram, including droplet crystal states in which a set of droplets arrange into a lattice pattern that breaks the rotational symmetry. An analytic model for small droplet crystals is developed and used to obtain a zero temperature phase diagram that is numerically validated. We show that in certain regimes a coherent low-density halo surrounds the droplet crystal giving rise to a novel phase with localized and extended features. 
\end{abstract} 

\maketitle 
Quantum droplets occur in ultra-cold atomic gases when quantum fluctuations (QFs) stabilize the system against collapse due to attractive two-body interactions. Such droplets have now been studied using dipolar \cite{Kadau2016a,Ferrier-Barbut2016a,Chomaz2016a,Ferrier-Barbut2018a,Schmitt2016a} and spinor \cite{Cabrera2018a,Semeghini2018a} Bose-Einstein condensates.  For the dipolar system the long ranged and anisotropic dipole-dipole interactions (DDIs) cause the droplets to elongate into filaments and neighboring droplets to repel each other, potentially stabilizing multi-droplet configurations.

In dipolar experiments droplets are usually produced by reducing the $s$-wave scattering length until the condensate is dominated by DDIs and  collapses into one or more droplets. There is a rich interplay between the trap potential and the DDIs that can be used to control the behavior of dipolar gases \cite{Koch2008a,Chomaz2018a}, particularly in the production of droplets: In prolate traps a single large droplet forms \cite{Chomaz2016a,Schmitt2016a}, while in an oblate trap (more tightly confined along the dipole direction) an array of small droplets emerges \cite{Kadau2016a}. It has been shown \cite{Blakie2016a,Wachtler2016a,Bisset2016a} that a single droplet is the ground state for the oblate trap studied in experiments \cite{Kadau2016a}, but because the condensate-droplet transition is discontinuous (also see \cite{Ferrier-Barbut2018a}) the system is unable to follow the ground state as the scattering length is ramped and many small droplets nucleate heating the system. Thus the observed multi-droplet states have high entropy and the droplets are not mutually coherent (see \cite{Bisset2015a,Xi2016a,Wenzel2017a}).
 \begin{figure}[!htbp] 
    \centering 
      \includegraphics[width=3.5in]{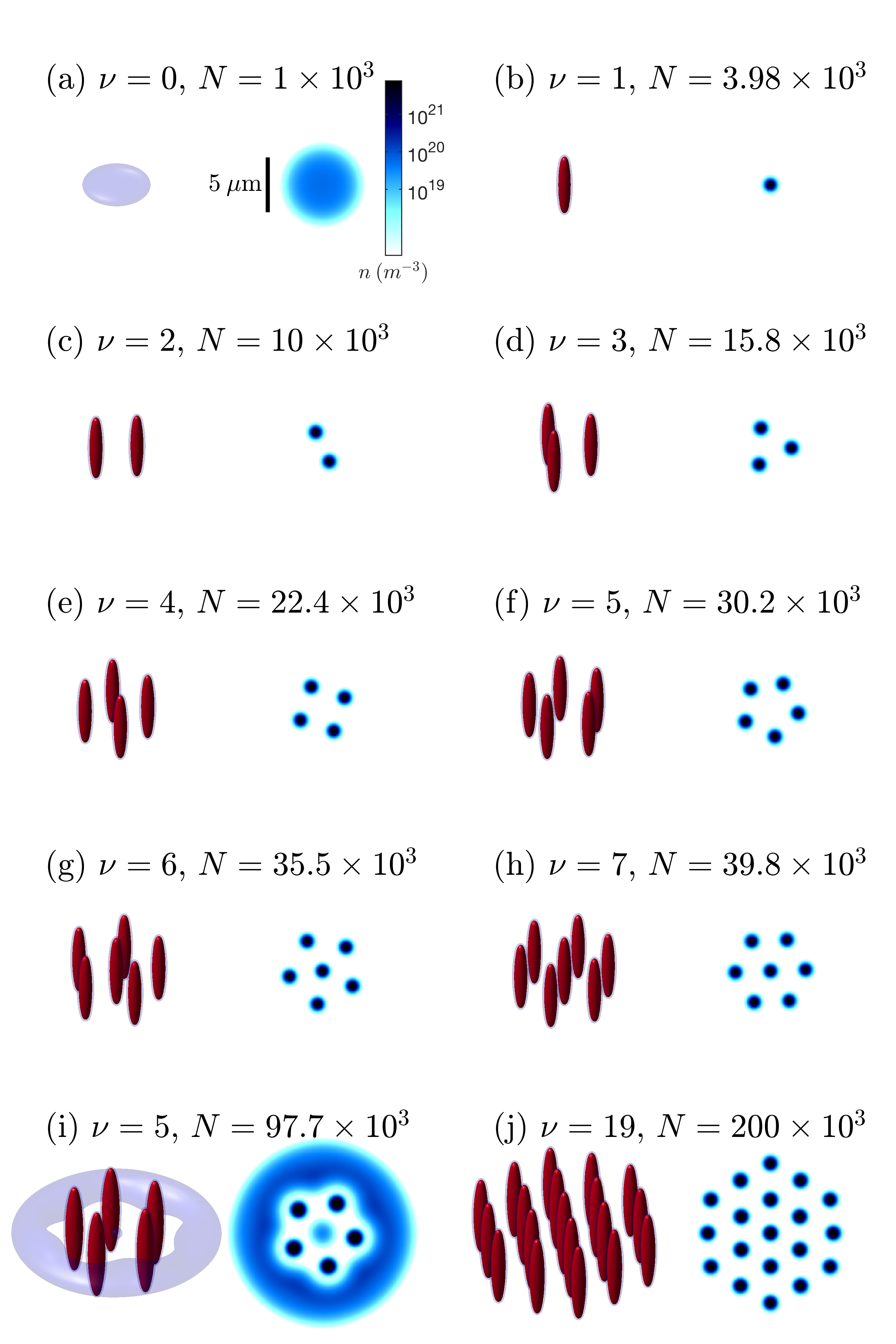}  
       \caption{  
   Stationary solutions of the extended GPE (\ref{e:GPE}) for a $^{164}$Dy gas with $a_s=70a_0$ in a $(\omega,\omega_z)=2\pi\times(60,300)$ Hz trap, atom numbers and  $\nu$ values (see text) as indicated.
   Left subplots:  Density isosurfaces at $\SI{2e19}{\per\metre\cubed}$ (blue) and  $\SI{2e20}{\per\metre\cubed}$ (red).
Right subplots: $z=0$ density slices.
   \label{f:nudroplets}}
\end{figure}
 
This poses the question as to whether droplet crystals can occur as the ground state of a dipolar gas in a regime where phase coherence persists across the droplets? In this case the ground state is a supersolid, a phase of matter than has recently been observed in condensates coupled to optical fields \cite{Leonard2017a,Li2017a}.  
  Here we address this question by exploring the ground state configurations of an oblately trapped dipolar condensate. Our key finding is that droplet crystals can occur as the ground state and low lying excited states of the system in parameter regimes accessible to current experiments (e.g.~see Fig.~\ref{f:nudroplets}). We study the energetics of the droplet crystal states and develop an analytic model to describe crystals, with explicit solutions given for up to 7 droplets. This allows us to determine a phase diagram for the crystal configurations which we validate against full numerical calculations of the extended Gross-Pitaevskii equation (GPE). Finally we demonstrate the emergence of haloed states in droplet crystals with sufficiently high chemical potential. In these states the droplets coexist with a low density orbital that typically extends around the outside of the crystal. Our results demonstrate that dipolar gases are an exciting dilute atomic system for studying supersolidity (cf.~\cite{Lu2015a,Cinti2010a,Cinti2017a}). We also note recent work predicting a striped phase of a dipolar condensate in a cigar shaped trap potential \cite{Wenzel2017a}.

\emph{Formalism}-- 
The system of interest is a dilute Bose gas of atoms with a magnetic moment $\mu_m$ polarized along $z$.
The stationary states of this system are described by the extended GPE  $\mu\psi=\LGP\psi$, where 
\begin{equation}
\LGP\psi=\left[ -\frac{\hbar^2 \nabla^2}{2m} + V(\bx) + \Phi(\bx) + \gammaQF |\psi|^3\right]\psi.\label{e:GPE}
\end{equation}
Here $\mu$ is the chemical potential and
\begin{equation}
    \Phi(\bx) = \int d\bx'\,U(\bx-\bx')|\psi(\bx')|^2, \label{e:Phix}
\end{equation}
with  $U(\br) = g_s\delta(\br) + \frac{3\gdd}{4\pi r^3}(1-3\frac{z^2}{r^2})$, where
 the contact interaction coupling constant is $g_s = 4\pi\hbar^2a_s/m$, $a_s$ is the $s$-wave scattering length,  and the dipolar interaction coupling constant is
$\gdd = \mu_0\mu_m^2/3$. 
 We define the dipole length $\add = m\gdd/4\pi\hbar^2$ and the ratio of dipolar to $s$-wave lengths $\edd = \gdd/g_s$. We include quantum fluctuations in the local density approximation, where the coefficient of this term is $\gammaQF = \frac{32}{3}g_s\sqrt{\frac{a_s^3}{\pi}}(1+\tfrac32 \edd^2)$ \cite{Lima2011a,Wachtler2016a,Bisset2016a}. The atoms are confined in a harmonic trapping potential $V=\frac{m}{2}\sum_{\alpha=x,y,z}\omega_\alpha^2\alpha^2$.
The extended GPE theory we have outlined is valid while the diluteness parameter $\sqrt{n\add^3}$ remains small, where $n$ is the number density, so that higher order quantum correlations are negligible.  This theory has successfully described droplet experiments with Er and Dy atoms (e.g.~see \cite{Ferrier-Barbut2016a,Schmitt2016a}), including detailed quantitative comparisons \cite{Chomaz2016a}. Here the densest  droplets shown have $n\add^3\lesssim 0.003$.

 Here we  focus on the case of a cylindrically symmetric trap with an oblate geometry with respect to the dipole polarization direction (i.e.~$\omega\equiv\omega_x=\omega_y$, with $\omega<\omega_z)$, as this configuration is conducive to the droplets subdividing driven by the tighter $z$ confinement.

\emph{Numerical results}-- 
We numerically represent the field $\psi$ on a three-dimensional mesh of points and use fast Fourier transforms to evaluate the kinetic energy and the convolution in Eq.~(\ref{e:Phix}). To improve the accuracy of the $\Phi$ calculation, a cylindrically cutoff DDI potential in $k$-space is used (e.g.~see \cite{Lu2010a}). We solve for stationary states using gradient flow \cite{Bao2010a} and conjugate gradient optimization \cite{Antoine2017a,Ronen2006a} techniques adapted to the extended GPE and for constrained total atom number $N=\int d\mathbf{x}|\psi|^2$. Solutions are accepted as converged when the residual $\max|\LGP\psi-\mu\psi|/\sqrt{N}$ is smaller than $10^{-4}$.

\begin{figure}[htbp] 
    \centering
      \includegraphics[width=3.5in]{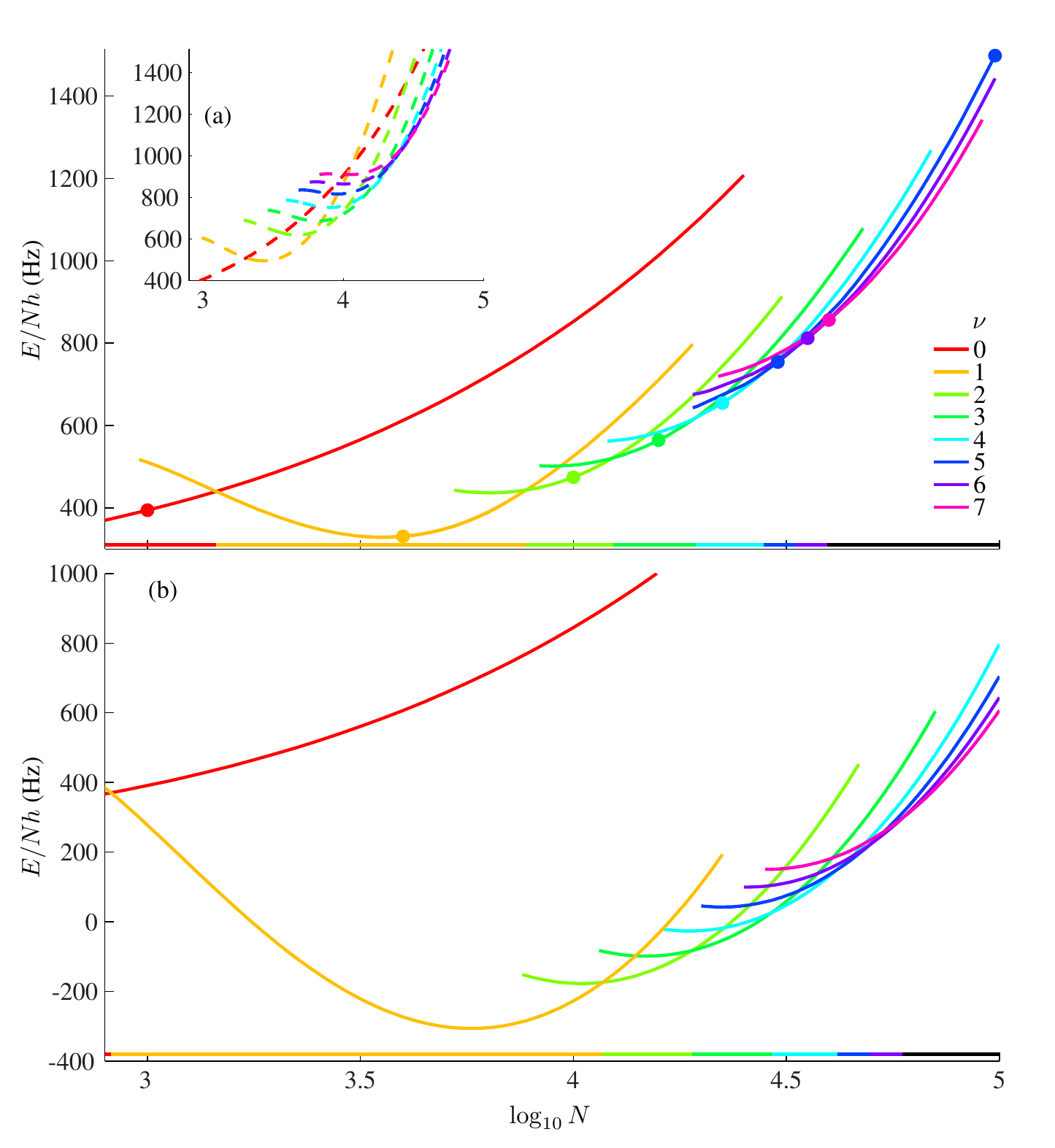} 
        \caption{ (a) Energy as a function of total atom number for $a_s=70a_0$ using the GPE solutions obtained the trap $(\omega,\omega_z) = 2\pi(60,300)$Hz.  Filled circle markers indicate the first nine states appearing in Fig.~\ref{f:nudroplets}.
      The inset shows the predictions of the variational solution. (b) As above but for  $a_s=65a_0$. Horizontal colored lines indicate the ground state at each $N$ value with arbitrary scale.  } 
   \label{f:Ecurves}
\end{figure}

In the regimes we study here, the system has a complicated energy landscape with many local minima corresponding to stationary states $\psi$ with different droplet crystal configurations. Individual solution runs starting from different initial fields or using different solvers can end up with qualitatively different stationary states.  We find that these states can be categorized as either a condensate or a particular type of droplet state. The condensate state is a low density solution in which the QF corrections are negligible, and for sufficiently large $N$, the density profile is determined by balancing the two-body interactions against the trapping potential. The droplet states only occur for $\edd>1$ and sufficiently many atoms, and consist of $\nu$ (positive integer) distinct dense filament shaped droplets that are elongated in the direction of the dipoles.  Examples of these different types of states are shown in Fig.~\ref{f:nudroplets}, where we have denoted the condensate state as $\nu=0$.

Since the numerical solvers we use can find various $\nu$-droplet states corresponding to local energy minima, we are able to follow particular states as the system parameters change to obtain the solution energy branches (e.g.~see Fig.~\ref{f:Ecurves}), where the energy is given by
\begin{equation}
E\! = \!\int \!d\bx \,\psi^*\!\left[ -\frac{\hbar^2 \nabla^2}{2m} \!+\! V(\bx) \!+ \!\frac12\Phi(\bx) \!+ \!\frac25 \gammaQF |\psi|^3\right]\!\psi.\label{e:Efunc}
\end{equation}
We can only follow a given branch over a limited parameter range until the state becomes dynamically or energetically unstable and decays to another (lower energy) branch. Excited stationary states are important since they are robust to perturbations and can be long lived. Indeed, excited multi-droplet arrays have been observed to be stable for $\gtrsim100$ ms \cite{Kadau2016a}. 

In Fig.~\ref{f:Ecurves}(a) we show the solution energies for $a_s=70a_0$ and $\nu\le7$ as a function of  atom number, showing that as $N$ increases the ground state transitions to crystals with an increasing number of droplets. Notably, the ground state changes from $\nu=0$ to $7$ as $N$ varies from   $1.4\times10^3$ to $4\times10^4$.

We can qualitatively explain the behavior of the energy branches in Fig.~\ref{f:Ecurves}(a). For the $\nu=0$ condensate solution $E/N$ increases with $N$. This is because the effective two-body interactions are repulsive as the DDI is tuned positive by the oblate condensate density profile (i.e.~oblate density enhances the repulsive side-by-side interaction of the dipoles).  In contrast for the droplet solutions the effective two-body interactions are attractive due to the prolate shape of the droplets. Following the single droplet ($\nu=1$) solution we see that $E/N$ initially decreases with increasing $N$. In this regime the droplet is small and the trap a plays negligible role, i.e.~it is effectively a self-bound droplet in free space \cite{Baillie2016b,Wachtler2016b}. 
The size of the droplet increases with $N$ and eventually the $z$-confinement becomes important. As this occurs we observe $E/N$ to level off and then start increasing. For sufficiently large $N$ it is favorable for the droplet to divide into two smaller droplets (with $N/2$ atoms each) to decrease the $z$-confinement energy (i.e.~when the $\nu=2$ curve descends below the $\nu=1$ curve at $N\approx8\times10^3$) \footnote{In free-space $E/N$ decreases with increasing $N$ for the single droplet and it is never energetically favorable to form several droplets.}. In the $\nu=2$ state, the droplets radially separate to positions where the repulsive DDI between the droplets is balanced by the radial trap forces.
As $N$ (and hence each droplet size) continues to increase it becomes favorable for states with more droplets to be the ground state. At higher $N$ where large $\nu$ values are preferred, the droplets spread out further in the trap  and the energy difference between higher $\nu$ branches tends to be smaller. We have not tracked $\nu>7$ solutions in detail, but note $\nu\ge7$ solutions become energetically favorable for the parameters of Fig.~\ref{f:Ecurves}(a) when $N>4\times10^4$. An example of a larger droplet crystal with $N=2\times10^5$ and $\nu=19$ is shown in Fig.~\ref{f:nudroplets}.

For a lower value of  $a_s$ (i.e.~higher $\edd$) the two-body interactions are more attractive in the droplet, allowing it to self-bind and become the ground state at lower atom number \cite{Baillie2016b}, see  Fig.~\ref{f:Ecurves}(b). Also, because the droplets are smaller and more dense, the transition to higher $\nu$ states occurs at larger $N$.

 \begin{figure}[htbp] 
    \centering 
      \includegraphics[width=3.5in]{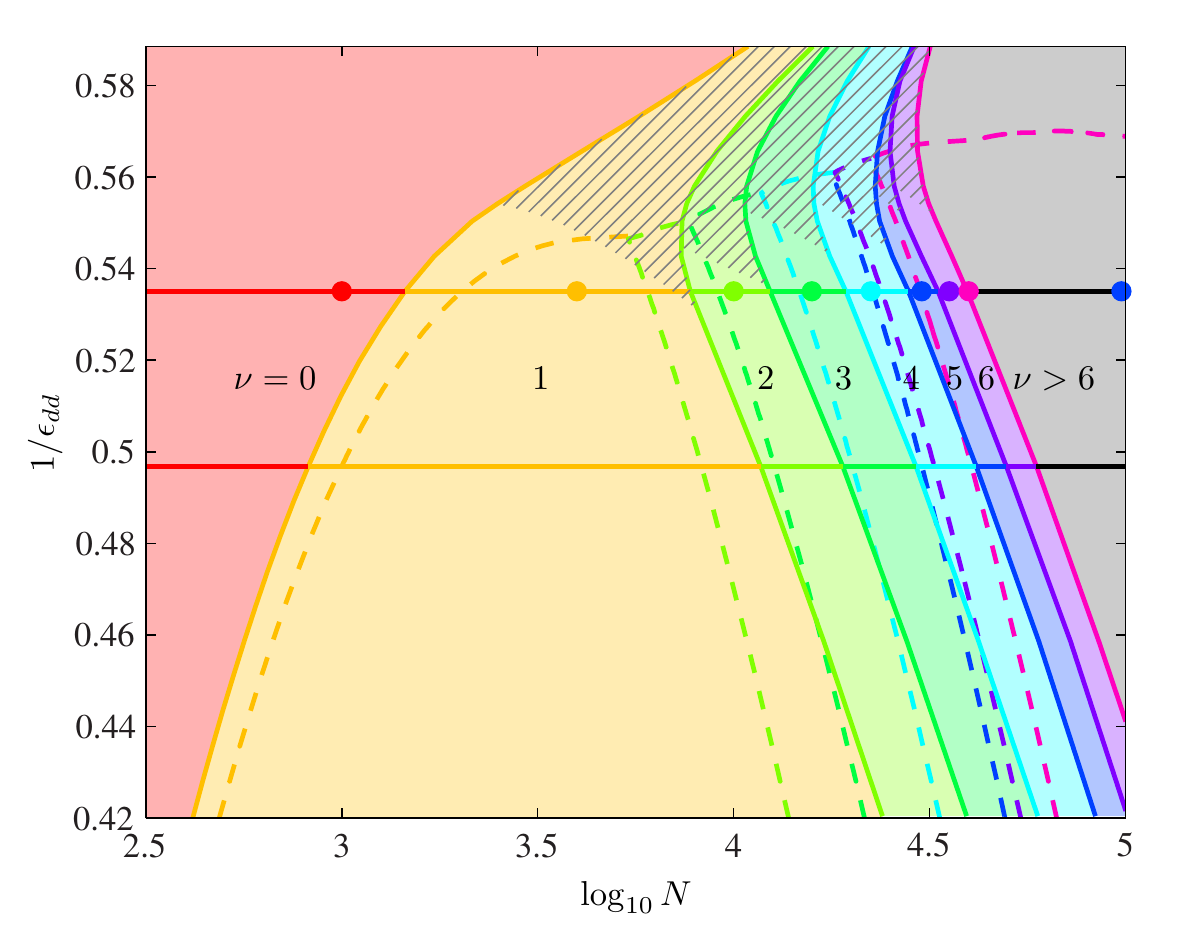}
      \caption{Ground state phase diagram  for $\nu$-droplet crystal states of  $^{164}$Dy atoms in a $(\omega,\omega_z) = 2\pi(60,300)$Hz trap. The solid lines and colored regions show the extended GPE results for where the $\nu$-droplet states with $\nu\le6$ are the lowest energy state. The gray region is where the ground state has $\nu>6$. Transition lines between states found using the variational model are indicated by dashed lines. The horizontal lines indicate the results of  Fig.~\ref{f:Ecurves} and the filled circles correspond to states in Fig.~\ref{f:nudroplets}(a)-(i). Hatched regions are where $ N_\mathrm{halo}/N> 1\%$.} 
   \label{f:phasediag}
\end{figure}

\emph{Variational model and phase diagram}-- We can formalize our qualitative discussion of droplet crystal energetics with a variational model.
We define an ansatz for $\nu\ge1$ droplets with total density $n(\bx) = \sum_{j=1}^\nu |\psi_\var(\bx-\bd_j)|^2$ where
\begin{equation}
    \psi_\var(\bx) = \sqrt{\frac{8N/\nu}{\pi^{3/2}\prod_\alpha\sigma_\alpha}} e^{-2\sum_\alpha \alpha^2/\sigma_\alpha^2},
\end{equation}
is the droplet orbital taken to be identical for all droplets, $\alpha=\{x,y,z\}$, and $\bd_j$ is the center of mass position for the $j$-th droplet.
Here $\bsigma$ and $\bd_j$ are variational parameters and we ignore overlap between the droplets. We calculate the intra-droplet dipolar energy using \eqref{e:Phix}, but for the inter-droplet dipolar energy we treat each droplet as a point dipole of $N/\nu$ atoms. We further assume that external trap confinement is sufficient so that the droplets positions $\bd_j$ lie in the $xy$-plane (as we find from the extended GPE calculations). %

In our ansatz the energy decouples into the sum of individual droplet energies which do not depend on $\bd_j$ and have been evaluated in \cite{Bisset2016a}, and a lattice energy
\begin{equation}
    E_L= \frac{m\omega^2 N}{2\nu} \sum_{j=1}^\nu |\bd_j|^2  + \frac{3\gdd N^2}{8\pi \nu^2 }\sum_{k\ne j}^\nu\frac{1}{|\bd_j-\bd_k|^3},\label{eq:EL}
\end{equation}
containing terms from the spatial arrangement of the droplets and the inter-droplet dipolar repulsion, which does not depend on $\bsigma$. Using the length scale $a_L= (3\add a_\omega^4 N/\nu)^{1/5}$ gives $E_L =m\omega^2a_L^2 \Eb_LN/2\nu$ where $a_\omega=\sqrt{\hbar/m\omega}$ and
\begin{equation}
    \Eb_L = \sum_{j=1}^\nu|\bdb_j|^2 + \sum_{k\ne j}^\nu\frac{1}{|\bdb_j-\bdb_k|^3}.\label{e:ELnu}
\end{equation}
For $\nu\le7$, $E_L$ is minimized when the droplets are located at the vertices of a regular polygon, but with one droplet at the center for $\nu=6,7$, i.e.~the number of vertices $\nu'$ is $\nu$ for $\nu<6$ and $\nu-1$ for $\nu=6,7$. The droplet positions are $\bdb_j = d_\nu \hat{d}_j$, with polygon radius $|\bd_j|=a_L d_\nu$ and $\hat{d}_j = [-\sin(\frac{2\pi j}{\nu'}\!+\!\phi),\cos(\frac{2\pi j}{\nu'}\!+\!\phi),0]$, for $j=0,\dots,\nu'$ and $\phi$ arbitrary. For $\nu=6,7$, $\hat{d}_\nu=\mathbf0$.  Then $\Eb_L=d_\nu^2\nu'  + d_\nu^{-3}s_\nu$ where $s_\nu = \sum_{j\ne k}^\nu |\hat{d}_j-\hat{d}_k|^{-3}$, which evaluates to $s_\nu = \{\frac14, \frac2{\sqrt3}, \frac12 + 2\sqrt2, 2\sqrt{2(5+\sqrt5)}, 2[5+\sqrt{2(5+\sqrt5)}], \frac{99}{4}+\frac4{\sqrt3}\vphantom{\dfrac1{\frac12}}\}$  for $\nu=2,\dots,7$.
The minimum of $\Eb_L$ is at $d_\nu = (\frac{3s_\nu}{2\nu'})^{1/5}\sim 1$
giving
\begin{equation}
    \frac{E_L}{N\hbar\omega} = \frac{5}{2\times 3^{3/5}} \frac{\nu'}{\nu}d_\nu^{\,2} \left(\frac{N\add}{\nu a_\omega}\right)^{2/5}. 
\end{equation}
Our full GPE solutions match these detailed predictions for the droplet crystal structure for $\nu\le7$ (e.g.~see Fig.~\ref{f:nudroplets}). 
For higher $\nu$, the droplets eventually form a triangular lattice [e.g.~Fig.~\ref{f:nudroplets}(j)] often with some distortion at the edges of the crystal. For comparison the total variational energy is shown in the inset to Fig.~\ref{f:Ecurves}(a).

In Fig.~\ref{f:phasediag} we present a phase diagram as a function of $N$ and $1/\edd$, showing the regions where the condensate and various droplet crystal configurations (up to $\nu=7$) are the ground states.  These results show that the variational theory provides a useful quantitative description of the full extended GPE solutions over a wide parameter regime.

\begin{figure}[htbp] 
   \centering
      \includegraphics[width=3.5in]{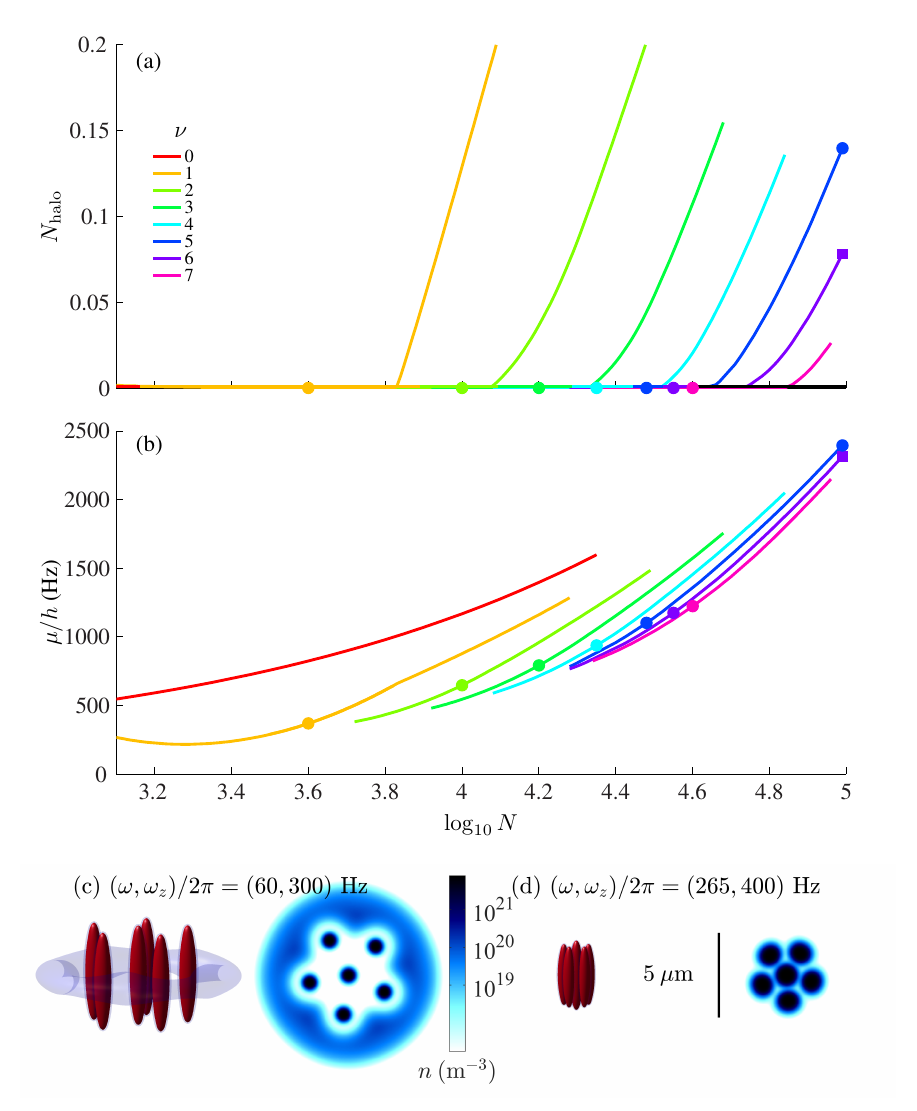}
      \caption{(a) Number of atoms in the halo $N_\mathrm{halo}$ as $N$ increases and (b) corresponding chemical potentials. Filled circles indicate the states (b)-(i) in Fig.~\ref{f:nudroplets}. Filled square indicates state (c) below. (c,d) $\nu=6$ droplet crystal states with $N=97.7\times10^3$. Same parameters as Fig.~\ref{f:Ecurves}(a), except for the higher trap frequencies in (d).} 
   \label{f:halo}  
\end{figure}
 
 \emph{Haloed droplet crystal states}-- 
As $N$ increases the droplet states develop a halo of low density atoms that extend outside the droplets [e.g.~see Figs.~\ref{f:nudroplets}(i), \ref{f:halo}(c)].   This occurs when the chemical potential is high enough for atoms to escape from the effective potential that self-binds the droplets, and spread out over accessible regions of the effective potential between the droplets. We quantify the number of atoms participating in the halo by calculating the normalization of $\psi$ excluding atoms within the high density droplets. The results in  Figs.~\ref{f:halo}(a-b) show that the halo suddenly forms at a critical number and chemical potential that increases with $\nu$. In the phase diagram Fig.~\ref{f:phasediag} we indicate where halo states occur. The variational model is not able to represent halo states, and in the regions of the phase diagram where these develop the variational predictions for the phase boundaries are  noticeably worse.

\emph{Conclusions and outlook}--
For the droplet crystals we have studied here to be a supersolid, they must be phase coherent \footnote{Note we  do not restrict our attention  to ground state crystals, as low energy excited states have long   lifetimes and favorable properties.}.  However, when the droplets are well separated, the tunneling between them is small and the coherence between droplets will vanish. To quantify this requires going beyond the classical field description underlying the extended GPE.
As an approximate treatment we can consider a pair of droplets in the crystal and model their relative coherence as a bosonic Josephson junction.
To do this we calculate the tunneling matrix element $J$ and charging energy $E_C$ from the stationary solutions \footnote{We take $J=\Delta E/2Nz_n$, where $\Delta E$ is energy change  when the phase of one droplet is twisted by $\pi$ and $z_n$ is the number of nearest neighbor droplets;  $E_C$ is evaluated as twice the second derivative of the energy (\ref{e:Efunc}) with respect of the normalization of the particular droplet orbital, see \cite{Giovanazzi2008a}.}. When $J>E_C$ (Josephson regime) there will be number fluctuations in each droplet and the state will be highly coherent with a well defined relative phase. When $J<E_C$ (Fock regime) 
the system prefers a well-defined atom number in each droplet, the coherence between them vanishes and the relative phase is undefined \cite{Gati2007a}.
As an example, for the central droplet of the $\nu=6$ state in Fig.~\ref{f:halo}(c) we find  $J/h=6\times 10^{-5}$ Hz and  $E_C/h=0.47$ Hz. In this case, where the droplets are $3.6\,\mu$m apart, the high relative cost of number fluctuations will suppress tunneling, and hence the droplets will be in the Fock regime and there will be no coherence across the droplet crystal. The tunneling can be enhanced by bringing the droplets closer together using tighter radial confinement \cite{Wenzel2017a}.  As an example, in Fig.~\ref{f:halo}(d) we show a $\nu=6$ state with tighter confinement, where the droplet separation has reduced to $1.5\,\mu$m. Here $J/h=2.2$ Hz and $E_C/h=0.38$ Hz, which is well within the Josephson regime, so coherence will extend between droplets.  
These results suggest the interesting possibility of observing a Mott insulator-like transition from  a supersolid to incoherent droplet crystal by slowly reducing the radial trap confinement.  We also note that the state shown in Fig.~\ref{f:halo}(c) has a halo, and while the droplets are mutually incoherent the halo part of the system will be coherent.

Future work will consider collective excitations of the droplet crystals (cf.~\cite{Baillie2017a}) and efficient experimental strategies for preparing particular crystal configurations.

\begin{acknowledgments}
We acknowledge the contribution of NZ eScience Infrastructure (NeSI) high-performance computing facilities, support from the Marsden Fund of the Royal Society of New Zealand, and valuable discussions with F.~Ferlaino and L.~Chomaz.
\end{acknowledgments}


\begin{thebibliography}{33}%
\makeatletter
\providecommand \@ifxundefined [1]{%
 \@ifx{#1\undefined}
}%
\providecommand \@ifnum [1]{%
 \ifnum #1\expandafter \@firstoftwo
 \else \expandafter \@secondoftwo
 \fi
}%
\providecommand \@ifx [1]{%
 \ifx #1\expandafter \@firstoftwo
 \else \expandafter \@secondoftwo
 \fi
}%
\providecommand \natexlab [1]{#1}%
\providecommand \enquote  [1]{``#1''}%
\providecommand \bibnamefont  [1]{#1}%
\providecommand \bibfnamefont [1]{#1}%
\providecommand \citenamefont [1]{#1}%
\providecommand \href@noop [0]{\@secondoftwo}%
\providecommand \href [0]{\begingroup \@sanitize@url \@href}%
\providecommand \@href[1]{\@@startlink{#1}\@@href}%
\providecommand \@@href[1]{\endgroup#1\@@endlink}%
\providecommand \@sanitize@url [0]{\catcode `\\12\catcode `\$12\catcode
  `\&12\catcode `\#12\catcode `\^12\catcode `\_12\catcode `\%12\relax}%
\providecommand \@@startlink[1]{}%
\providecommand \@@endlink[0]{}%
\providecommand \url  [0]{\begingroup\@sanitize@url \@url }%
\providecommand \@url [1]{\endgroup\@href {#1}{\urlprefix }}%
\providecommand \urlprefix  [0]{URL }%
\providecommand \Eprint [0]{\href }%
\providecommand \doibase [0]{http://dx.doi.org/}%
\providecommand \selectlanguage [0]{\@gobble}%
\providecommand \bibinfo  [0]{\@secondoftwo}%
\providecommand \bibfield  [0]{\@secondoftwo}%
\providecommand \translation [1]{[#1]}%
\providecommand \BibitemOpen [0]{}%
\providecommand \bibitemStop [0]{}%
\providecommand \bibitemNoStop [0]{.\EOS\space}%
\providecommand \EOS [0]{\spacefactor3000\relax}%
\providecommand \BibitemShut  [1]{\csname bibitem#1\endcsname}%
\let\auto@bib@innerbib\@empty
%</preamble>
\bibitem [{\citenamefont {Kadau}\ \emph {et~al.}(2016)\citenamefont {Kadau},
  \citenamefont {Schmitt}, \citenamefont {Wenzel}, \citenamefont {Wink},
  \citenamefont {Maier}, \citenamefont {Ferrier-Barbut},\ and\ \citenamefont
  {Pfau}}]{Kadau2016a}%
  \BibitemOpen
  \bibfield  {author} {\bibinfo {author} {\bibfnamefont {H.}~\bibnamefont
  {Kadau}}, \bibinfo {author} {\bibfnamefont {M.}~\bibnamefont {Schmitt}},
  \bibinfo {author} {\bibfnamefont {M.}~\bibnamefont {Wenzel}}, \bibinfo
  {author} {\bibfnamefont {C.}~\bibnamefont {Wink}}, \bibinfo {author}
  {\bibfnamefont {T.}~\bibnamefont {Maier}}, \bibinfo {author} {\bibfnamefont
  {I.}~\bibnamefont {Ferrier-Barbut}}, \ and\ \bibinfo {author} {\bibfnamefont
  {T.}~\bibnamefont {Pfau}},\ }\bibfield  {title} {\enquote {\bibinfo {title}
  {Observing the {R}osensweig instability of a quantum ferrofluid},}\ }\href
  {http://dx.doi.org/10.1038/nature16485} {\bibfield  {journal} {\bibinfo
  {journal} {Nature}\ }\textbf {\bibinfo {volume} {530}},\ \bibinfo {pages}
  {194} (\bibinfo {year} {2016})}\BibitemShut {NoStop}%
\bibitem [{\citenamefont {Ferrier-Barbut}\ \emph {et~al.}(2016)\citenamefont
  {Ferrier-Barbut}, \citenamefont {Kadau}, \citenamefont {Schmitt},
  \citenamefont {Wenzel},\ and\ \citenamefont {Pfau}}]{Ferrier-Barbut2016a}%
  \BibitemOpen
  \bibfield  {author} {\bibinfo {author} {\bibfnamefont {I.}~\bibnamefont
  {Ferrier-Barbut}}, \bibinfo {author} {\bibfnamefont {H.}~\bibnamefont
  {Kadau}}, \bibinfo {author} {\bibfnamefont {M.}~\bibnamefont {Schmitt}},
  \bibinfo {author} {\bibfnamefont {M.}~\bibnamefont {Wenzel}}, \ and\ \bibinfo
  {author} {\bibfnamefont {T.}~\bibnamefont {Pfau}},\ }\bibfield  {title}
  {\enquote {\bibinfo {title} {Observation of quantum droplets in a strongly
  dipolar {B}ose gas},}\ }\href {\doibase 10.1103/PhysRevLett.116.215301}
  {\bibfield  {journal} {\bibinfo  {journal} {Phys. Rev. Lett.}\ }\textbf
  {\bibinfo {volume} {116}},\ \bibinfo {pages} {215301} (\bibinfo {year}
  {2016})}\BibitemShut {NoStop}%
\bibitem [{\citenamefont {Chomaz}\ \emph {et~al.}(2016)\citenamefont {Chomaz},
  \citenamefont {Baier}, \citenamefont {Petter}, \citenamefont {Mark},
  \citenamefont {W\"achtler}, \citenamefont {Santos},\ and\ \citenamefont
  {Ferlaino}}]{Chomaz2016a}%
  \BibitemOpen
  \bibfield  {author} {\bibinfo {author} {\bibfnamefont {L.}~\bibnamefont
  {Chomaz}}, \bibinfo {author} {\bibfnamefont {S.}~\bibnamefont {Baier}},
  \bibinfo {author} {\bibfnamefont {D.}~\bibnamefont {Petter}}, \bibinfo
  {author} {\bibfnamefont {M.~J.}\ \bibnamefont {Mark}}, \bibinfo {author}
  {\bibfnamefont {F.}~\bibnamefont {W\"achtler}}, \bibinfo {author}
  {\bibfnamefont {L.}~\bibnamefont {Santos}}, \ and\ \bibinfo {author}
  {\bibfnamefont {F.}~\bibnamefont {Ferlaino}},\ }\bibfield  {title} {\enquote
  {\bibinfo {title} {Quantum-fluctuation-driven crossover from a dilute
  {B}ose-{E}instein condensate to a macrodroplet in a dipolar quantum fluid},}\
  }\href {\doibase 10.1103/PhysRevX.6.041039} {\bibfield  {journal} {\bibinfo
  {journal} {Phys. Rev. X}\ }\textbf {\bibinfo {volume} {6}},\ \bibinfo {pages}
  {041039} (\bibinfo {year} {2016})}\BibitemShut {NoStop}%
\bibitem [{\citenamefont {Ferrier-Barbut}\ \emph {et~al.}(2018)\citenamefont
  {Ferrier-Barbut}, \citenamefont {Wenzel}, \citenamefont {Schmitt},
  \citenamefont {B\"ottcher},\ and\ \citenamefont
  {Pfau}}]{Ferrier-Barbut2018a}%
  \BibitemOpen
  \bibfield  {author} {\bibinfo {author} {\bibfnamefont {I.}~\bibnamefont
  {Ferrier-Barbut}}, \bibinfo {author} {\bibfnamefont {M.}~\bibnamefont
  {Wenzel}}, \bibinfo {author} {\bibfnamefont {M.}~\bibnamefont {Schmitt}},
  \bibinfo {author} {\bibfnamefont {F.}~\bibnamefont {B\"ottcher}}, \ and\
  \bibinfo {author} {\bibfnamefont {T.}~\bibnamefont {Pfau}},\ }\bibfield
  {title} {\enquote {\bibinfo {title} {Onset of a modulational instability in
  trapped dipolar {B}ose-{E}instein condensates},}\ }\href {\doibase %
  10.1103/PhysRevA.97.011604} {\bibfield  {journal} {\bibinfo  {journal} {Phys.
  Rev. A}\ }\textbf {\bibinfo {volume} {97}},\ \bibinfo {pages} {011604(R)}
  (\bibinfo {year} {2018})}\BibitemShut {NoStop}%
\bibitem [{\citenamefont {Schmitt}\ \emph {et~al.}(2016)\citenamefont
  {Schmitt}, \citenamefont {Wenzel}, \citenamefont {B{\"o}ttcher},
  \citenamefont {Ferrier-Barbut},\ and\ \citenamefont {Pfau}}]{Schmitt2016a}%
  \BibitemOpen
  \bibfield  {author} {\bibinfo {author} {\bibfnamefont {M.}~\bibnamefont
  {Schmitt}}, \bibinfo {author} {\bibfnamefont {M.}~\bibnamefont {Wenzel}},
  \bibinfo {author} {\bibfnamefont {F.}~\bibnamefont {B{\"o}ttcher}}, \bibinfo
  {author} {\bibfnamefont {I.}~\bibnamefont {Ferrier-Barbut}}, \ and\ \bibinfo
  {author} {\bibfnamefont {T.}~\bibnamefont {Pfau}},\ }\bibfield  {title}
  {\enquote {\bibinfo {title} {Self-bound droplets of a dilute magnetic quantum
  liquid},}\ }\href {http://dx.doi.org/10.1038/nature20126} {\bibfield
  {journal} {\bibinfo  {journal} {Nature}\ }\textbf {\bibinfo {volume} {539}},\
  \bibinfo {pages} {259} (\bibinfo {year} {2016})}\BibitemShut {NoStop}%
\bibitem [{\citenamefont {Cabrera}\ \emph {et~al.}(2018)\citenamefont
  {Cabrera}, \citenamefont {Tanzi}, \citenamefont {Sanz}, \citenamefont
  {Naylor}, \citenamefont {Thomas}, \citenamefont {Cheiney},\ and\
  \citenamefont {Tarruell}}]{Cabrera2018a}%
  \BibitemOpen
  \bibfield  {author} {\bibinfo {author} {\bibfnamefont {C.~R.}\ \bibnamefont
  {Cabrera}}, \bibinfo {author} {\bibfnamefont {L.}~\bibnamefont {Tanzi}},
  \bibinfo {author} {\bibfnamefont {J.}~\bibnamefont {Sanz}}, \bibinfo {author}
  {\bibfnamefont {B.}~\bibnamefont {Naylor}}, \bibinfo {author} {\bibfnamefont
  {P.}~\bibnamefont {Thomas}}, \bibinfo {author} {\bibfnamefont
  {P.}~\bibnamefont {Cheiney}}, \ and\ \bibinfo {author} {\bibfnamefont
  {L.}~\bibnamefont {Tarruell}},\ }\bibfield  {title} {\enquote {\bibinfo
  {title} {Quantum liquid droplets in a mixture of {B}ose-{E}instein
  condensates},}\ }\href {\doibase 10.1126/science.aao5686} {\bibfield
  {journal} {\bibinfo  {journal} {Science}\ }\textbf {\bibinfo {volume}
  {359}},\ \bibinfo {pages} {301} (\bibinfo {year} {2018})}\BibitemShut
  {NoStop}%
\bibitem [{\citenamefont {Semeghini}\ \emph {et~al.}(2018)\citenamefont
  {Semeghini}, \citenamefont {Ferioli}, \citenamefont {Masi}, \citenamefont
  {Mazzinghi}, \citenamefont {Wolswijk}, \citenamefont {Minardi}, \citenamefont
  {Modugno}, \citenamefont {Modugno}, \citenamefont {Inguscio},\ and\
  \citenamefont {Fattori}}]{Semeghini2018a}%
  \BibitemOpen
  \bibfield  {author} {\bibinfo {author} {\bibfnamefont {G.}~\bibnamefont
  {Semeghini}}, \bibinfo {author} {\bibfnamefont {G.}~\bibnamefont {Ferioli}},
  \bibinfo {author} {\bibfnamefont {L.}~\bibnamefont {Masi}}, \bibinfo {author}
  {\bibfnamefont {C.}~\bibnamefont {Mazzinghi}}, \bibinfo {author}
  {\bibfnamefont {L.}~\bibnamefont {Wolswijk}}, \bibinfo {author}
  {\bibfnamefont {F.}~\bibnamefont {Minardi}}, \bibinfo {author} {\bibfnamefont
  {M.}~\bibnamefont {Modugno}}, \bibinfo {author} {\bibfnamefont
  {G.}~\bibnamefont {Modugno}}, \bibinfo {author} {\bibfnamefont
  {M.}~\bibnamefont {Inguscio}}, \ and\ \bibinfo {author} {\bibfnamefont
  {M.}~\bibnamefont {Fattori}},\ }\bibfield  {title} {\enquote {\bibinfo
  {title} {Self-bound quantum droplets of atomic mixtures in free space},}\
  }\href {\doibase 10.1103/PhysRevLett.120.235301} {\bibfield  {journal}
  {\bibinfo  {journal} {Phys. Rev. Lett.}\ }\textbf {\bibinfo {volume} {120}},\
  \bibinfo {pages} {235301} (\bibinfo {year} {2018})}\BibitemShut {NoStop}%
\bibitem [{\citenamefont {Koch}\ \emph {et~al.}(2008)\citenamefont {Koch},
  \citenamefont {Lahaye}, \citenamefont {Metz}, \citenamefont {Frohlich},
  \citenamefont {Griesmaier},\ and\ \citenamefont {Pfau}}]{Koch2008a}%
  \BibitemOpen
  \bibfield  {author} {\bibinfo {author} {\bibfnamefont {T.}~\bibnamefont
  {Koch}}, \bibinfo {author} {\bibfnamefont {T.}~\bibnamefont {Lahaye}},
  \bibinfo {author} {\bibfnamefont {J.}~\bibnamefont {Metz}}, \bibinfo {author}
  {\bibfnamefont {B.}~\bibnamefont {Frohlich}}, \bibinfo {author}
  {\bibfnamefont {A.}~\bibnamefont {Griesmaier}}, \ and\ \bibinfo {author}
  {\bibfnamefont {T.}~\bibnamefont {Pfau}},\ }\bibfield  {title} {\enquote
  {\bibinfo {title} {Stabilization of a purely dipolar quantum gas against
  collapse},}\ }\href {\doibase 10.1038/nphys887} {\bibfield  {journal}
  {\bibinfo  {journal} {Nat. Phys.}\ }\textbf {\bibinfo {volume} {4}},\
  \bibinfo {pages} {218} (\bibinfo {year} {2008})}\BibitemShut {NoStop}%
\bibitem [{\citenamefont {Chomaz}\ \emph {et~al.}(2018)\citenamefont {Chomaz},
  \citenamefont {van Bijnen}, \citenamefont {Petter}, \citenamefont {Faraoni},
  \citenamefont {Baier}, \citenamefont {Becher}, \citenamefont {Mark},
  \citenamefont {W{\"a}chtler}, \citenamefont {Santos},\ and\ \citenamefont
  {Ferlaino}}]{Chomaz2018a}%
  \BibitemOpen
  \bibfield  {author} {\bibinfo {author} {\bibfnamefont {L.}~\bibnamefont
  {Chomaz}}, \bibinfo {author} {\bibfnamefont {R.~M.~W.}\ \bibnamefont {van
  Bijnen}}, \bibinfo {author} {\bibfnamefont {D.}~\bibnamefont {Petter}},
  \bibinfo {author} {\bibfnamefont {G.}~\bibnamefont {Faraoni}}, \bibinfo
  {author} {\bibfnamefont {S.}~\bibnamefont {Baier}}, \bibinfo {author}
  {\bibfnamefont {J.~H.}\ \bibnamefont {Becher}}, \bibinfo {author}
  {\bibfnamefont {M.~J.}\ \bibnamefont {Mark}}, \bibinfo {author}
  {\bibfnamefont {F.}~\bibnamefont {W{\"a}chtler}}, \bibinfo {author}
  {\bibfnamefont {L.}~\bibnamefont {Santos}}, \ and\ \bibinfo {author}
  {\bibfnamefont {F.}~\bibnamefont {Ferlaino}},\ }\bibfield  {title} {\enquote
  {\bibinfo {title} {Observation of roton mode population in a dipolar quantum
  gas},}\ }\href {\doibase 10.1038/s41567-018-0054-7} {\bibfield  {journal}
  {\bibinfo  {journal} {Nat. Phys.}\ }\textbf {\bibinfo {volume} {14}},\
  \bibinfo {pages} {442} (\bibinfo {year} {2018})}\BibitemShut {NoStop}%
\bibitem [{\citenamefont {Blakie}(2016)}]{Blakie2016a}%
  \BibitemOpen
  \bibfield  {author} {\bibinfo {author} {\bibfnamefont {P.~B.}\ \bibnamefont
  {Blakie}},\ }\bibfield  {title} {\enquote {\bibinfo {title} {Properties of a
  dipolar condensate with three-body interactions},}\ }\href {\doibase %
  10.1103/PhysRevA.93.033644} {\bibfield  {journal} {\bibinfo  {journal} {Phys.
  Rev. A}\ }\textbf {\bibinfo {volume} {93}},\ \bibinfo {pages} {033644}
  (\bibinfo {year} {2016})}\BibitemShut {NoStop}%
\bibitem [{\citenamefont {W\"achtler}\ and\ \citenamefont
  {Santos}(2016{\natexlab{a}})}]{Wachtler2016a}%
  \BibitemOpen
  \bibfield  {author} {\bibinfo {author} {\bibfnamefont {F.}~\bibnamefont
  {W\"achtler}}\ and\ \bibinfo {author} {\bibfnamefont {L.}~\bibnamefont
  {Santos}},\ }\bibfield  {title} {\enquote {\bibinfo {title} {Quantum
  filaments in dipolar {B}ose-{E}instein condensates},}\ }\href {\doibase %
  10.1103/PhysRevA.93.061603} {\bibfield  {journal} {\bibinfo  {journal} {Phys.
  Rev. A}\ }\textbf {\bibinfo {volume} {93}},\ \bibinfo {pages} {061603}
  (\bibinfo {year} {2016}{\natexlab{a}})}\BibitemShut {NoStop}%
\bibitem [{\citenamefont {Bisset}\ \emph {et~al.}(2016)\citenamefont {Bisset},
  \citenamefont {Wilson}, \citenamefont {Baillie},\ and\ \citenamefont
  {Blakie}}]{Bisset2016a}%
  \BibitemOpen
  \bibfield  {author} {\bibinfo {author} {\bibfnamefont {R.~N.}\ \bibnamefont
  {Bisset}}, \bibinfo {author} {\bibfnamefont {R.~M.}\ \bibnamefont {Wilson}},
  \bibinfo {author} {\bibfnamefont {D.}~\bibnamefont {Baillie}}, \ and\
  \bibinfo {author} {\bibfnamefont {P.~B.}\ \bibnamefont {Blakie}},\ }\bibfield
   {title} {\enquote {\bibinfo {title} {Ground-state phase diagram of a dipolar
  condensate with quantum fluctuations},}\ }\href {\doibase %
  10.1103/PhysRevA.94.033619} {\bibfield  {journal} {\bibinfo  {journal} {Phys.
  Rev. A}\ }\textbf {\bibinfo {volume} {94}},\ \bibinfo {pages} {033619}
  (\bibinfo {year} {2016})}\BibitemShut {NoStop}%
\bibitem [{\citenamefont {Bisset}\ and\ \citenamefont
  {Blakie}(2015)}]{Bisset2015a}%
  \BibitemOpen
  \bibfield  {author} {\bibinfo {author} {\bibfnamefont {R.~N.}\ \bibnamefont
  {Bisset}}\ and\ \bibinfo {author} {\bibfnamefont {P.~B.}\ \bibnamefont
  {Blakie}},\ }\bibfield  {title} {\enquote {\bibinfo {title} {Crystallization
  of a dilute atomic dipolar condensate},}\ }\href {\doibase %
  10.1103/PhysRevA.92.061603} {\bibfield  {journal} {\bibinfo  {journal} {Phys.
  Rev. A}\ }\textbf {\bibinfo {volume} {92}},\ \bibinfo {pages} {061603}
  (\bibinfo {year} {2015})}\BibitemShut {NoStop}%
\bibitem [{\citenamefont {Xi}\ and\ \citenamefont {Saito}(2016)}]{Xi2016a}%
  \BibitemOpen
  \bibfield  {author} {\bibinfo {author} {\bibfnamefont {K.-T.}\ \bibnamefont
  {Xi}}\ and\ \bibinfo {author} {\bibfnamefont {H.}~\bibnamefont {Saito}},\
  }\bibfield  {title} {\enquote {\bibinfo {title} {Droplet formation in a
  {B}ose-{E}instein condensate with strong dipole-dipole interaction},}\ }\href
  {\doibase 10.1103/PhysRevA.93.011604} {\bibfield  {journal} {\bibinfo
  {journal} {Phys. Rev. A}\ }\textbf {\bibinfo {volume} {93}},\ \bibinfo
  {pages} {011604} (\bibinfo {year} {2016})}\BibitemShut {NoStop}%
\bibitem [{\citenamefont {Wenzel}\ \emph {et~al.}(2017)\citenamefont {Wenzel},
  \citenamefont {B\"ottcher}, \citenamefont {Langen}, \citenamefont
  {Ferrier-Barbut},\ and\ \citenamefont {Pfau}}]{Wenzel2017a}%
  \BibitemOpen
  \bibfield  {author} {\bibinfo {author} {\bibfnamefont {M.}~\bibnamefont
  {Wenzel}}, \bibinfo {author} {\bibfnamefont {F.}~\bibnamefont {B\"ottcher}},
  \bibinfo {author} {\bibfnamefont {T.}~\bibnamefont {Langen}}, \bibinfo
  {author} {\bibfnamefont {I.}~\bibnamefont {Ferrier-Barbut}}, \ and\ \bibinfo
  {author} {\bibfnamefont {T.}~\bibnamefont {Pfau}},\ }\bibfield  {title}
  {\enquote {\bibinfo {title} {Striped states in a many-body system of tilted
  dipoles},}\ }\href {\doibase 10.1103/PhysRevA.96.053630} {\bibfield
  {journal} {\bibinfo  {journal} {Phys. Rev. A}\ }\textbf {\bibinfo {volume}
  {96}},\ \bibinfo {pages} {053630} (\bibinfo {year} {2017})}\BibitemShut
  {NoStop}%
\bibitem [{\citenamefont {L{\'e}onard}\ \emph {et~al.}(2017)\citenamefont
  {L{\'e}onard}, \citenamefont {Morales}, \citenamefont {Zupancic},
  \citenamefont {Esslinger},\ and\ \citenamefont {Donner}}]{Leonard2017a}%
  \BibitemOpen
  \bibfield  {author} {\bibinfo {author} {\bibfnamefont {J.}~\bibnamefont
  {L{\'e}onard}}, \bibinfo {author} {\bibfnamefont {A.}~\bibnamefont
  {Morales}}, \bibinfo {author} {\bibfnamefont {P.}~\bibnamefont {Zupancic}},
  \bibinfo {author} {\bibfnamefont {T.}~\bibnamefont {Esslinger}}, \ and\
  \bibinfo {author} {\bibfnamefont {T.}~\bibnamefont {Donner}},\ }\bibfield
  {title} {\enquote {\bibinfo {title} {Supersolid formation in a quantum gas
  breaking a continuous translational symmetry},}\ }\href
  {http://dx.doi.org/10.1038/nature21067} {\bibfield  {journal} {\bibinfo
  {journal} {Nature}\ }\textbf {\bibinfo {volume} {543}},\ \bibinfo {pages}
  {87} (\bibinfo {year} {2017})}\BibitemShut {NoStop}%
\bibitem [{\citenamefont {Li}\ \emph {et~al.}(2017)\citenamefont {Li},
  \citenamefont {Lee}, \citenamefont {Huang}, \citenamefont {Burchesky},
  \citenamefont {Shteynas}, \citenamefont {Top}, \citenamefont {Jamison},\ and\
  \citenamefont {Ketterle}}]{Li2017a}%
  \BibitemOpen
  \bibfield  {author} {\bibinfo {author} {\bibfnamefont {J.-R.}\ \bibnamefont
  {Li}}, \bibinfo {author} {\bibfnamefont {J.}~\bibnamefont {Lee}}, \bibinfo
  {author} {\bibfnamefont {W.}~\bibnamefont {Huang}}, \bibinfo {author}
  {\bibfnamefont {S.}~\bibnamefont {Burchesky}}, \bibinfo {author}
  {\bibfnamefont {B.}~\bibnamefont {Shteynas}}, \bibinfo {author}
  {\bibfnamefont {F.~{\c{C}}.}\ \bibnamefont {Top}}, \bibinfo {author}
  {\bibfnamefont {A.~O.}\ \bibnamefont {Jamison}}, \ and\ \bibinfo {author}
  {\bibfnamefont {W.}~\bibnamefont {Ketterle}},\ }\bibfield  {title} {\enquote
  {\bibinfo {title} {A stripe phase with supersolid properties in
  spin--orbit-coupled {B}ose--{E}instein condensates},}\ }\href
  {http://dx.doi.org/10.1038/nature21431} {\bibfield  {journal} {\bibinfo
  {journal} {Nature}\ }\textbf {\bibinfo {volume} {543}},\ \bibinfo {pages}
  {91} (\bibinfo {year} {2017})}\BibitemShut {NoStop}%
\bibitem [{\citenamefont {Lu}\ \emph {et~al.}(2015)\citenamefont {Lu},
  \citenamefont {Li}, \citenamefont {Petrov},\ and\ \citenamefont
  {Shlyapnikov}}]{Lu2015a}%
  \BibitemOpen
  \bibfield  {author} {\bibinfo {author} {\bibfnamefont {Z.-K.}\ \bibnamefont
  {Lu}}, \bibinfo {author} {\bibfnamefont {Y.}~\bibnamefont {Li}}, \bibinfo
  {author} {\bibfnamefont {D.~S.}\ \bibnamefont {Petrov}}, \ and\ \bibinfo
  {author} {\bibfnamefont {G.~V.}\ \bibnamefont {Shlyapnikov}},\ }\bibfield
  {title} {\enquote {\bibinfo {title} {Stable dilute supersolid of
  two-dimensional dipolar bosons},}\ }\href {\doibase %
  10.1103/PhysRevLett.115.075303} {\bibfield  {journal} {\bibinfo  {journal}
  {Phys. Rev. Lett.}\ }\textbf {\bibinfo {volume} {115}},\ \bibinfo {pages}
  {075303} (\bibinfo {year} {2015})}\BibitemShut {NoStop}%
\bibitem [{\citenamefont {Cinti}\ \emph {et~al.}(2010)\citenamefont {Cinti},
  \citenamefont {Jain}, \citenamefont {Boninsegni}, \citenamefont {Micheli},
  \citenamefont {Zoller},\ and\ \citenamefont {Pupillo}}]{Cinti2010a}%
  \BibitemOpen
  \bibfield  {author} {\bibinfo {author} {\bibfnamefont {F.}~\bibnamefont
  {Cinti}}, \bibinfo {author} {\bibfnamefont {P.}~\bibnamefont {Jain}},
  \bibinfo {author} {\bibfnamefont {M.}~\bibnamefont {Boninsegni}}, \bibinfo
  {author} {\bibfnamefont {A.}~\bibnamefont {Micheli}}, \bibinfo {author}
  {\bibfnamefont {P.}~\bibnamefont {Zoller}}, \ and\ \bibinfo {author}
  {\bibfnamefont {G.}~\bibnamefont {Pupillo}},\ }\bibfield  {title} {\enquote
  {\bibinfo {title} {Supersolid droplet crystal in a dipole-blockaded gas},}\
  }\href {\doibase 10.1103/PhysRevLett.105.135301} {\bibfield  {journal}
  {\bibinfo  {journal} {Phys. Rev. Lett.}\ }\textbf {\bibinfo {volume} {105}},\
  \bibinfo {pages} {135301} (\bibinfo {year} {2010})}\BibitemShut {NoStop}%
\bibitem [{\citenamefont {Cinti}\ and\ \citenamefont
  {Boninsegni}(2017)}]{Cinti2017a}%
  \BibitemOpen
  \bibfield  {author} {\bibinfo {author} {\bibfnamefont {F.}~\bibnamefont
  {Cinti}}\ and\ \bibinfo {author} {\bibfnamefont {M.}~\bibnamefont
  {Boninsegni}},\ }\bibfield  {title} {\enquote {\bibinfo {title} {Classical
  and quantum filaments in the ground state of trapped dipolar {Bose} gases},}\
  }\href {\doibase 10.1103/PhysRevA.96.013627} {\bibfield  {journal} {\bibinfo
  {journal} {Phys. Rev. A}\ }\textbf {\bibinfo {volume} {96}},\ \bibinfo
  {pages} {013627} (\bibinfo {year} {2017})}\BibitemShut {NoStop}%
\bibitem [{\citenamefont {Lima}\ and\ \citenamefont
  {Pelster}(2011)}]{Lima2011a}%
  \BibitemOpen
  \bibfield  {author} {\bibinfo {author} {\bibfnamefont {A.~R.~P.}\
  \bibnamefont {Lima}}\ and\ \bibinfo {author} {\bibfnamefont {A.}~\bibnamefont
  {Pelster}},\ }\bibfield  {title} {\enquote {\bibinfo {title} {Quantum
  fluctuations in dipolar {B}ose gases},}\ }\href {\doibase %
  10.1103/PhysRevA.84.041604} {\bibfield  {journal} {\bibinfo  {journal} {Phys.
  Rev. A}\ }\textbf {\bibinfo {volume} {84}},\ \bibinfo {pages} {041604}
  (\bibinfo {year} {2011})}\BibitemShut {NoStop}%
\bibitem [{\citenamefont {Lu}\ \emph {et~al.}(2010)\citenamefont {Lu},
  \citenamefont {Lu}, \citenamefont {Zhang}, \citenamefont {Qiu}, \citenamefont
  {Pu},\ and\ \citenamefont {Yi}}]{Lu2010a}%
  \BibitemOpen
  \bibfield  {author} {\bibinfo {author} {\bibfnamefont {H.-Y.}\ \bibnamefont
  {Lu}}, \bibinfo {author} {\bibfnamefont {H.}~\bibnamefont {Lu}}, \bibinfo
  {author} {\bibfnamefont {J.-N.}\ \bibnamefont {Zhang}}, \bibinfo {author}
  {\bibfnamefont {R.-Z.}\ \bibnamefont {Qiu}}, \bibinfo {author} {\bibfnamefont
  {H.}~\bibnamefont {Pu}}, \ and\ \bibinfo {author} {\bibfnamefont
  {S.}~\bibnamefont {Yi}},\ }\bibfield  {title} {\enquote {\bibinfo {title}
  {Spatial density oscillations in trapped dipolar condensates},}\ }\href
  {\doibase 10.1103/PhysRevA.82.023622} {\bibfield  {journal} {\bibinfo
  {journal} {Phys. Rev. A}\ }\textbf {\bibinfo {volume} {82}},\ \bibinfo
  {pages} {023622} (\bibinfo {year} {2010})}\BibitemShut {NoStop}%
\bibitem [{\citenamefont {Bao}\ \emph {et~al.}(2010)\citenamefont {Bao},
  \citenamefont {Cai},\ and\ \citenamefont {Wang}}]{Bao2010a}%
  \BibitemOpen
  \bibfield  {author} {\bibinfo {author} {\bibfnamefont {W.}~\bibnamefont
  {Bao}}, \bibinfo {author} {\bibfnamefont {Y.}~\bibnamefont {Cai}}, \ and\
  \bibinfo {author} {\bibfnamefont {H.}~\bibnamefont {Wang}},\ }\bibfield
  {title} {\enquote {\bibinfo {title} {Efficient numerical methods for
  computing ground states and dynamics of dipolar {B}ose–{E}instein
  condensates},}\ }\href {\doibase 10.1016/j.jcp.2010.07.001} {\bibfield
  {journal} {\bibinfo  {journal} {J. Comput. Phys.}\ }\textbf {\bibinfo
  {volume} {229}},\ \bibinfo {pages} {7874 } (\bibinfo {year}
  {2010})}\BibitemShut {NoStop}%
\bibitem [{\citenamefont {Antoine}\ \emph {et~al.}(2017)\citenamefont
  {Antoine}, \citenamefont {Levitt},\ and\ \citenamefont
  {Tang}}]{Antoine2017a}%
  \BibitemOpen
  \bibfield  {author} {\bibinfo {author} {\bibfnamefont {X.}~\bibnamefont
  {Antoine}}, \bibinfo {author} {\bibfnamefont {A.}~\bibnamefont {Levitt}}, \
  and\ \bibinfo {author} {\bibfnamefont {Q.}~\bibnamefont {Tang}},\ }\bibfield
  {title} {\enquote {\bibinfo {title} {Efficient spectral computation of the
  stationary states of rotating {B}ose–{E}instein condensates by
  preconditioned nonlinear conjugate gradient methods},}\ }\href {\doibase %
  10.1016/j.jcp.2017.04.040} {\bibfield  {journal} {\bibinfo  {journal} {J.
  Comput. Phys.}\ }\textbf {\bibinfo {volume} {343}},\ \bibinfo {pages} {92 }
  (\bibinfo {year} {2017})}\BibitemShut {NoStop}%
\bibitem [{\citenamefont {Ronen}\ \emph {et~al.}(2006)\citenamefont {Ronen},
  \citenamefont {Bortolotti},\ and\ \citenamefont {Bohn}}]{Ronen2006a}%
  \BibitemOpen
  \bibfield  {author} {\bibinfo {author} {\bibfnamefont {S.}~\bibnamefont
  {Ronen}}, \bibinfo {author} {\bibfnamefont {D.~C.~E.}\ \bibnamefont
  {Bortolotti}}, \ and\ \bibinfo {author} {\bibfnamefont {J.~L.}\ \bibnamefont
  {Bohn}},\ }\bibfield  {title} {\enquote {\bibinfo {title} {{B}ogoliubov modes
  of a dipolar condensate in a cylindrical trap},}\ }\href {\doibase %
  10.1103/PhysRevA.74.013623} {\bibfield  {journal} {\bibinfo  {journal} {Phys.
  Rev. A}\ }\textbf {\bibinfo {volume} {74}},\ \bibinfo {pages} {013623}
  (\bibinfo {year} {2006})}\BibitemShut {NoStop}%
\bibitem [{\citenamefont {Baillie}\ \emph {et~al.}(2016)\citenamefont
  {Baillie}, \citenamefont {Wilson}, \citenamefont {Bisset},\ and\
  \citenamefont {Blakie}}]{Baillie2016b}%
  \BibitemOpen
  \bibfield  {author} {\bibinfo {author} {\bibfnamefont {D.}~\bibnamefont
  {Baillie}}, \bibinfo {author} {\bibfnamefont {R.~M.}\ \bibnamefont {Wilson}},
  \bibinfo {author} {\bibfnamefont {R.~N.}\ \bibnamefont {Bisset}}, \ and\
  \bibinfo {author} {\bibfnamefont {P.~B.}\ \bibnamefont {Blakie}},\ }\bibfield
   {title} {\enquote {\bibinfo {title} {Self-bound dipolar droplet: A localized
  matter wave in free space},}\ }\href {\doibase 10.1103/PhysRevA.94.021602}
  {\bibfield  {journal} {\bibinfo  {journal} {Phys. Rev. A}\ }\textbf {\bibinfo
  {volume} {94}},\ \bibinfo {pages} {021602(R)} (\bibinfo {year}
  {2016})}\BibitemShut {NoStop}%
\bibitem [{\citenamefont {W\"achtler}\ and\ \citenamefont
  {Santos}(2016{\natexlab{b}})}]{Wachtler2016b}%
  \BibitemOpen
  \bibfield  {author} {\bibinfo {author} {\bibfnamefont {F.}~\bibnamefont
  {W\"achtler}}\ and\ \bibinfo {author} {\bibfnamefont {L.}~\bibnamefont
  {Santos}},\ }\bibfield  {title} {\enquote {\bibinfo {title} {Ground-state
  properties and elementary excitations of quantum droplets in dipolar
  {B}ose-{E}instein condensates},}\ }\href {\doibase %
  10.1103/PhysRevA.94.043618} {\bibfield  {journal} {\bibinfo  {journal} {Phys.
  Rev. A}\ }\textbf {\bibinfo {volume} {94}},\ \bibinfo {pages} {043618}
  (\bibinfo {year} {2016}{\natexlab{b}})}\BibitemShut {NoStop}%
\bibitem [{Note1()}]{Note1}%
  \BibitemOpen
  \bibinfo {note} {In free-space $E/N$ decreases with increasing $N$ for the
  single droplet and it is never energetically favorable to form several
  droplets.}\BibitemShut {Stop}%
\bibitem [{Note2()}]{Note2}%
  \BibitemOpen
  \bibinfo {note} {Note we do not restrict our attention to ground state
  crystals, as low energy excited states have long lifetimes and favorable
  properties.}\BibitemShut {Stop}%
\bibitem [{Note3()}]{Note3}%
  \BibitemOpen
  \bibinfo {note} {We take $J=\Delta E/2Nz_n$, where $\Delta E$ is energy
  change when the phase of one droplet is twisted by $\pi $ and $z_n$ is the
  number of nearest neighbor droplets; $E_C$ is evaluated as twice the second
  derivative of the energy (\ref {e:Efunc}) with respect of the normalization
  of the particular droplet orbital, see \cite {Giovanazzi2008a}.}\BibitemShut
  {Stop}%
\bibitem [{\citenamefont {Gati}\ and\ \citenamefont
  {Oberthaler}(2007)}]{Gati2007a}%
  \BibitemOpen
  \bibfield  {author} {\bibinfo {author} {\bibfnamefont {R.}~\bibnamefont
  {Gati}}\ and\ \bibinfo {author} {\bibfnamefont {M.~K.}\ \bibnamefont
  {Oberthaler}},\ }\bibfield  {title} {\enquote {\bibinfo {title} {A bosonic
  {J}osephson junction},}\ }\href
  {http://stacks.iop.org/0953-4075/40/i=10/a=R01} {\bibfield  {journal}
  {\bibinfo  {journal} {J. Phys. B}\ }\textbf {\bibinfo {volume} {40}},\
  \bibinfo {pages} {R61} (\bibinfo {year} {2007})}\BibitemShut {NoStop}%
\bibitem [{\citenamefont {Baillie}\ \emph {et~al.}(2017)\citenamefont
  {Baillie}, \citenamefont {Wilson},\ and\ \citenamefont
  {Blakie}}]{Baillie2017a}%
  \BibitemOpen
  \bibfield  {author} {\bibinfo {author} {\bibfnamefont {D.}~\bibnamefont
  {Baillie}}, \bibinfo {author} {\bibfnamefont {R.~M.}\ \bibnamefont {Wilson}},
  \ and\ \bibinfo {author} {\bibfnamefont {P.~B.}\ \bibnamefont {Blakie}},\
  }\bibfield  {title} {\enquote {\bibinfo {title} {Collective excitations of
  self-bound droplets of a dipolar quantum fluid},}\ }\href {\doibase %
  10.1103/PhysRevLett.119.255302} {\bibfield  {journal} {\bibinfo  {journal}
  {Phys. Rev. Lett.}\ }\textbf {\bibinfo {volume} {119}},\ \bibinfo {pages}
  {255302} (\bibinfo {year} {2017})}\BibitemShut {NoStop}%
\bibitem [{\citenamefont {Giovanazzi}\ \emph {et~al.}(2008)\citenamefont
  {Giovanazzi}, \citenamefont {Esteve},\ and\ \citenamefont
  {Oberthaler}}]{Giovanazzi2008a}%
  \BibitemOpen
  \bibfield  {author} {\bibinfo {author} {\bibfnamefont {S.}~\bibnamefont
  {Giovanazzi}}, \bibinfo {author} {\bibfnamefont {J.}~\bibnamefont {Esteve}},
  \ and\ \bibinfo {author} {\bibfnamefont {M.~K.}\ \bibnamefont {Oberthaler}},\
  }\bibfield  {title} {\enquote {\bibinfo {title} {Effective parameters for
  weakly coupled {B}ose-{E}instein condensates},}\ }\href
  {http://stacks.iop.org/1367-2630/10/i=4/a=045009} {\bibfield  {journal}
  {\bibinfo  {journal} {New J. Phys.}\ }\textbf {\bibinfo {volume} {10}},\
  \bibinfo {pages} {045009} (\bibinfo {year} {2008})}\BibitemShut {NoStop}%
\end{thebibliography}
\end{document}